\documentclass[aps,prc,twocolumn,showpacs,floatfix,nofootinbib,preprintnumbers,superscriptaddress,amsmath,amssymb]{revtex4-1}

\usepackage{epsfig}
\usepackage{color}
\usepackage{natbib}
\usepackage{graphicx}
\usepackage{xifthen}
\usepackage[percent]{overpic}

\newcommand{\al}{$\alpha$}
\newcommand{\Carb}[1][]{\ifthenelse{\equal{#1}{}}{$^{12}$C}{$^{#1}$C}}
\newcommand{\Be}[1][]{\ifthenelse{\equal{#1}{}}{$^{8}$Be}{$^{#1}$Be}}
\newcommand{\Ox}[1][]{\ifthenelse{\equal{#1}{}}{$^{16}$O}{$^{#1}$O}}
\newcommand{\Ca}[1][]{\ifthenelse{\equal{#1}{}}{$^{40}$Ca}{$^{#1}$Ca}}

\newcommand{\loc}[1][]{\ifthenelse{\equal{#1}{}}{\ensuremath{\mathcal{C}}}{\ensuremath{\mathcal{C}_{#1}}}}

\newcommand{\fmd}{\,\mathrm{fm^{-3}}}
\newcommand{\fmt}{\,\mathrm{fm/c}}
\newcommand{\fm}{\,\mathrm{fm}}
\newcommand{\MeV}{\,\mathrm{MeV}}
\newcommand{\Ecm}{E_\mathrm{cm}}

\renewcommand{\vec}[1]{\mbox{\boldmath $#1$}}

\begin{document}

\title{Cluster formation in pre-compound nuclei in the time-dependent framework}

\author{B. Schuetrumpf}
\affiliation{Institut f\"ur Kerphysik, Technische Universit\"at Darmstadt, Schlossgartenstra\ss{}e 2, 64289 Darmstadt, Germany}
\affiliation{GSI Helmholzzentrum f\"ur Schwerionenforschung, Planckstra\ss{}e 1, 64291 Darmstadt, Germany}
\affiliation{FRIB Laboratory, Michigan State University, East Lansing, Michigan 48824, USA}

\author{W. Nazarewicz}
\affiliation{Department of Physics and Astronomy and FRIB Laboratory, Michigan State University, East Lansing, Michigan 48824, USA}

\date{\today}

\begin{abstract}
\begin{description}
\item[Background]
Modern applications of nuclear time-dependent density functional theory (TDDFT)  are often capable of providing quantitative description of heavy ion reactions. However, the structure of pre-compound (pre-equilibrium, pre-fission) states produced in heavy ion reactions are difficult to assess theoretically in TDDFT as the single-particle density alone  is a weak indicator of shell structure and cluster states.
\item[Purpose]
We employ the time-dependent nucleon localization function (NLF)  to reveal the structure of pre-compound states in nuclear reactions involving light and medium-mass ions. We primarily focus  on spin saturated systems with $N=Z$. Furthermore, we study reactions with oxygen and carbon ions, for which some experimental evidence for \al{} clustering in  pre-compound states exists.
\item[Method]
We utilize the symmetry-free TDDFT approach with the Skyrme energy density functional UNEDF1 and compute the time-dependent NLFs to  describe \Ox{} + \Ox{}, \Ca{} + \Ox{}, \Ca{} + \Ca{}, and \Ox[16,18]{} + \Carb{} collisions at energies  above the Coulomb barrier.
\item[Results]
We show that NLFs reveal a variety of time-dependent modes involving cluster structures. For instance, the \Ox{} + \Ox{} collision results in a vibrational mode of a quasi-molecular \al{}-\Carb{}-\Carb{}-\al{}  state. For heavier ions, a variety of  cluster configurations are predicted. 
For the collision of \Ox[16,18]{} + \Carb{}, we showed that the pre-compound system has a tendency to form \al{} clusters. This result
supports the experimental findings that the presence of
cluster structures in the  projectile and target nuclei gives rise to
strong entrance channel effects and enhanced \al{} emission.
\item[Conclusion]
The time-dependent nucleon localization measure is a very good indicator of clusters structures  in  complex pre-compound states formed in heavy-ion fusion reactions. The localization reveals the presence of collective  vibrations involving cluster structures,  which dominate the initial dynamics of the fusing system.
\end{description}
\end{abstract}

\maketitle
\section{Introduction}

Low-energy fusion initiated by light and medium-mass ions is of great importance for both basic science and applications. In many cases, this process can be well described in terms of the compound nucleus framework, which assumes that the
excited composite nucleus formed in the fusion reaction
lives long enough for the thermodynamic equilibrium to be established. 
In many cases, however, entrance channel effects can be significant, and an idealized picture of  a  compound nucleus, which  is not expected to retain  memory of how it was formed, is clearly not appropriate \cite{Birkelund,Lebhertz,Nagashima,Vadas15}.

In this work, we carry out theoretical study of entrance channel effects 
in low-energy  collisions of light heavy-ions using the time-dependent density functional theory (TDDFT) approach. In particular, we are interested in structure of nuclear configurations formed shortly following fusion. Those are
pre-compound (or pre-equilibrium or pre-fission)  nuclear states formed in fusion or fusion-fission reactions that carry significant memory of the entrance channel.

The self-consistent time-dependent Hartree-Fock theory, or its TDDFT extension, is a standard tool to study heavy-ion collisions  (see Refs.~\cite{Rei03a,Furche2005,Mar04,Marques06,Nakatsukasa16} for reviews). Advanced  symmetry-unrestricted TDDFT calculations for nuclear reactions and other large-amplitude collective motion, have been dramatically advanced by the use of high-performance  
computing
\cite{Umar1986,Kim97,Maruhn2005,Umar2005,Umar2006,Umar15,Umar16,Nakatsukasa2005,Simenel04,Simenel12,Simenel13,Sekizawa16,Schuetrumpf2016}. A useful extension of TDDFT is density constrained TDHF \cite{Umar2009,Oberacker10}, which can be used to extract  capture cross sections, including the sub-barrier regime. 

The structure of pre-compound states formed in fusion reactions can be impacted by the presence of clustering effects. Indeed, clustering has been shown to be very important theoretically in low-energy \cite{Bauer87} and high energy heavy-ion collisions \cite{Aichelin88,Zhang17}. Moreover, there exists some experimental evidence, as well as  quite a few  theoretical predictions,  for the presence of nuclear molecular states in light and medium-mass nuclei 
at high excitation energy \cite{Betts97,Beck00,Beck13,Kun99,Belyaeva,Ichikawa11}.

To study clustering effects in heavy-ion fusion reactions, we utilize a measure called fermion localization function, originally developed for electronic calculations \cite{Becke1990}. As demonstrated in Refs.~\cite{Reinhard2011,Zhang2016,Schuetrumpf2017}, the nucleon localization function (NLF) is an excellent tool to reveal shell and cluster effects in nuclei. In this work, we apply the concept of nuclear localization to analyze the structure of  states formed in heavy-ion collision TDDFT simulations. In this way, we can quantify the nature of pre-equilibrium configurations whose  structure is largely undiagnosed when using conventional techniques. 

This article is organized as follows. Section.~\ref{sec:theory} contains a brief description of the TDDFT formalism used in this work and describes nucleon localization functions. In Sec.~\ref{sec:sym} we analyze the symmetric collisions of the doubly magic nuclei  \Ox{} + \Ox{} and \Ca{} + \Ca{}, while in Sec.~\ref{sec:asym} we discuss the asymmetric collisions of \Ox{} + \Ca{} and \Ox[16,18]{} + \Carb{}. Finally, the summary and outlook are provided in  Sec.~\ref{sec:Conclusion}.

\section{Theoretical framework} \label{sec:theory}
\subsection{Nuclear TDDFT}
For the  TDDFT calculations we utilize the software package Sky3D \cite{Maruhn2014}, which solves the time-dependent Hartree-Fock equations in  the coordinate space on an equidistant grid using fast Fourier transforms (FFT) for derivatives. For the nuclear mean-field, we use the Skyrme functional UNEDF1
\cite{Kortelainen2012}, which is expected to perform well at large deformations.
UNEDF1 has been optimized to selected  properties of nuclei and nuclear matter and no additional parameters have been  introduced for the time-dependent calculations presented in this work. 

The initial wave functions of  colliding ions are determined from ground-state (g.s.) DFT calculations performed with Sky3D. For closed-shell systems pairing correlations are ignored. For the open-shell systems  \Ox[18]{} and \Carb, we use the BCS pairing as in Ref.~\cite{Bender2000}.  In the time-dependent calculations we use the frozen occupation approximation.

The wave functions of two colliding ions are combined into one  Slater determinant by orthogonalizing all single-particle wave functions. The fragments are boosted by multiplying the wave functions with a complex phase factor, and then evolved in time with a finite time step of $0.2\fmt$. 
The Sky3D framework does not impose any symmetry restrictions. However, while the long-range Coulomb problem is solved for open boundary conditions, the short-range nuclear interaction is determined in the box with periodic boundary conditions, since the FFT approach is used for computing derivatives. We took
a cubic  box with a large length of $32\fm$  to ensure that the wave functions vanish at the boundaries. Finite-volume effects can be practically eliminated  using the twist-averaged boundary condition  \cite{Schuetrumpf2015,Schuetrumpf2016a}. However, due to the relatively small time-scales considered, such an approach was not needed in this work.

While TDDFT calculations can well reproduce certain observables such as the fusion, or capture, cross sections above the Coulomb barrier, there are obvious limitations to the theory, such as its inability to describe the motion of the system in the classically forbidden region, many-body dissipation, and fluctuations due to internally broken symmetries  \cite{Simenel2011}. In particular, the transition to the compound-nucleus phase cannot be described within TDDFT. In this study, therefore, we shall limit our investigations to the pre-compound configurations involving relatively short time scales.

\subsection{Nucleon localization function}

The electron localization function  was originally proposed to characterize chemical bonding in electronic systems \cite{Becke1990,savin1997,scemama2004,Kohout04,burnus2005,Poater}. Subsequently,  the nucleon localization function (NLF) was applied to atomic nuclei  to visualize cluster structures in light systems \cite{Reinhard2011,Schuetrumpf2017}.  The NLF is derived from the inverse of the conditional probability of finding a nucleon of isospin $q$ ($n$, or $p$) in the vicinity of another nucleon of the same isospin and  signature quantum number $\sigma$ ($=\uparrow$ or $\downarrow$), knowing for certainty that the latter particle is located at position $\vec{r}$. The NLF can entirely be expressed through the local DFT densities:
\begin{equation} 
\mathcal{C}_{q\sigma}(\vec{r})=\left[1+\left(\frac{\tau_{q\sigma}\rho_{q\sigma}-\frac{1}{4}|\vec{\nabla}\rho_{q\sigma}|^2-\vec{j}^2_{q\sigma}}{\rho_{q\sigma}\tau^\mathrm{TF}_{q\sigma}}\right)^2\right]^{-1},
\label{eq:localization}
\end{equation}
where $\rho_{q\sigma}$, $\tau_{q\sigma}$, $\vec{j}_{q\sigma}$, and $\vec{\nabla}\rho_{q\sigma}$ are the particle density, kinetic energy density, current density, and density gradient, respectively, and $\tau^\mathrm{TF}_{q\sigma}$ denotes the Thomas-Fermi kinetic energy. 

In Ref.~\cite{Reinhard2011} mostly $N=Z$ nuclei up to $A=20$ have been studied. It was demonstrated that   \al{} clusters tend to  appear at the tips of deformed nuclei, and that  \Carb{} clusters can be  revealed through characteristic  rings of enhanced NLF. Furthermore,  NLFs have also been studied for heavy nuclei  \cite{Zhang2016} to investigate fragment formation during nuclear fission, and also to  pasta phases in the inner crust of neutron stars \cite{Schuetrumpf2017}.

While in g.s. calculations for even-even nuclei the time-reversal symmetry is conserved and the current density $\vec{j}_{q\sigma}$ vanishes, it becomes an important ingredient in time-dependent calculations. Furthermore,  since in this work we are primarily interested in the localization of  neutrons and protons, and not in the signature content, in the following  we consider signature-average densities, such as   $\rho_{q}=(\rho_{q\uparrow}+\rho_{q\downarrow})/2$.

To study  \al{}-clusters or clusters of \al{}-conjugate nuclei in light-to-medium $N=Z$ systems with weak Coulomb forces, it is convenient to utilize the \al{}-NLF as introduced in Ref.~\cite{Reinhard2011}:
\begin{equation}
\loc[\alpha]=\sqrt{\loc[n]\loc[p]}.
\end{equation} 
The localization function takes generally values between 0 and 1. High values of NLF indicate that the probability of finding two particles (of the same type) close to each other is low.  Since the localization function (\ref{eq:localization}) is normalized to the Thomas-Fermi kinetic energy, the value of $\mathcal{C}=1/2$ corresponds to a limit of the homogeneous Fermi gas, in which the individual orbits are spatially delocalized.

\begin{figure}[htb]
\includegraphics[width=0.8\columnwidth]{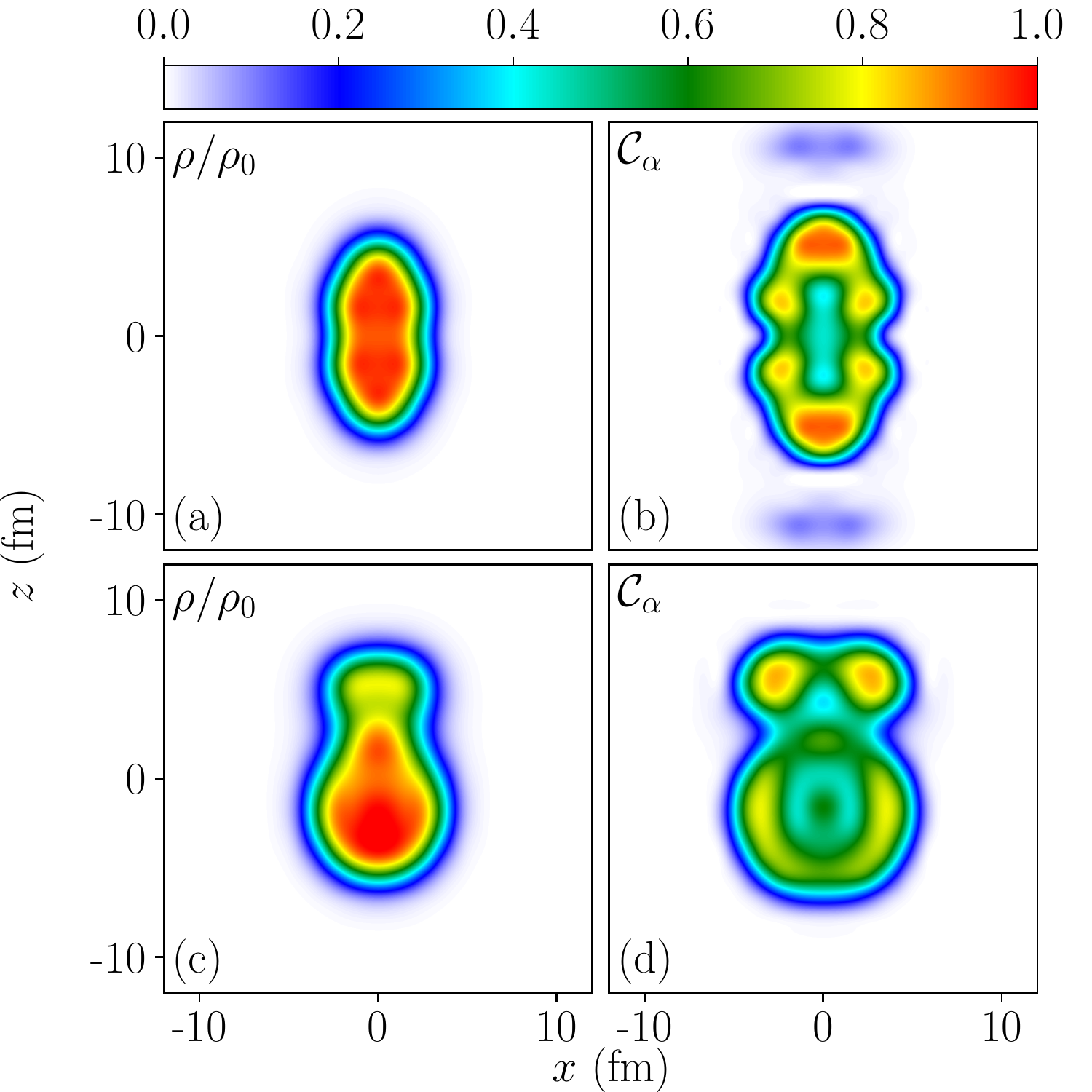}
\caption{\label{fig:rholoc} Snapshots of \Ox{} + \Ox{} 
(top) and \Ox{} + \Ca{} (bottom) TDDFT collision simulations. Total densities normalized to
the nuclear saturation density $\rho_0=0.16$\,fm$^{-3}$
are shown in the left panels while the corresponding localizations  $\loc[\alpha]$ are displayed in the right panels. 
Since the collisions are central, axial symmetry with respect to the z-axis is conserved.}
\end{figure}
Examples of the  density distribution and corresponding
localizations  $\loc[\alpha]$ predicted in TDDFT  are shown in Fig.~\ref{fig:rholoc}. In general, particle densities
contain little  information about the internal structure of the system.
On the other hand, the NLFs 
reveal  distinct regions with enhanced localization that signal the appearance of cluster structures. Regions where clusters overlap  exhibit decreased localization. In the following, we shall use localizations  $\loc[\alpha]$ 
to identify and visualize various cluster structures and their collective motion.

\subsection{Assessing nucleon content in clusters}

To complement the analysis based on NLFs, we extract the nucleon content in the spatial regions dominated by  single clusters. Such regions correspond to enhanced values of localization; they
are separated by areas  of $\loc[\alpha]\approx 0.5$ in which  the cluster wave functions overlap. For simplicity,  we only consider  central collisions and  assume that the clusters are located along the direction of the boost ($z$-axis). 
The nucleon content of a cluster identified by means of the NLF
is given by
\begin{equation}
A(z_1,z_2)=\iint dxdy \int_{z_1}^{z_2}\rho(x,y,z)dz.
\label{eq:integ_dens}
\end{equation}
The NLF offers some freedom to chose the  values of $z_1$ and $z_2$. Here,  we chose the values such that $A(z_1,z_2)$ is integer.


\section{Localization in symmetric collisions} \label{sec:sym}
\subsection{\Ox{} + \Ox{} collisions}

\begin{figure}[htb]
\includegraphics[width=.9\columnwidth]{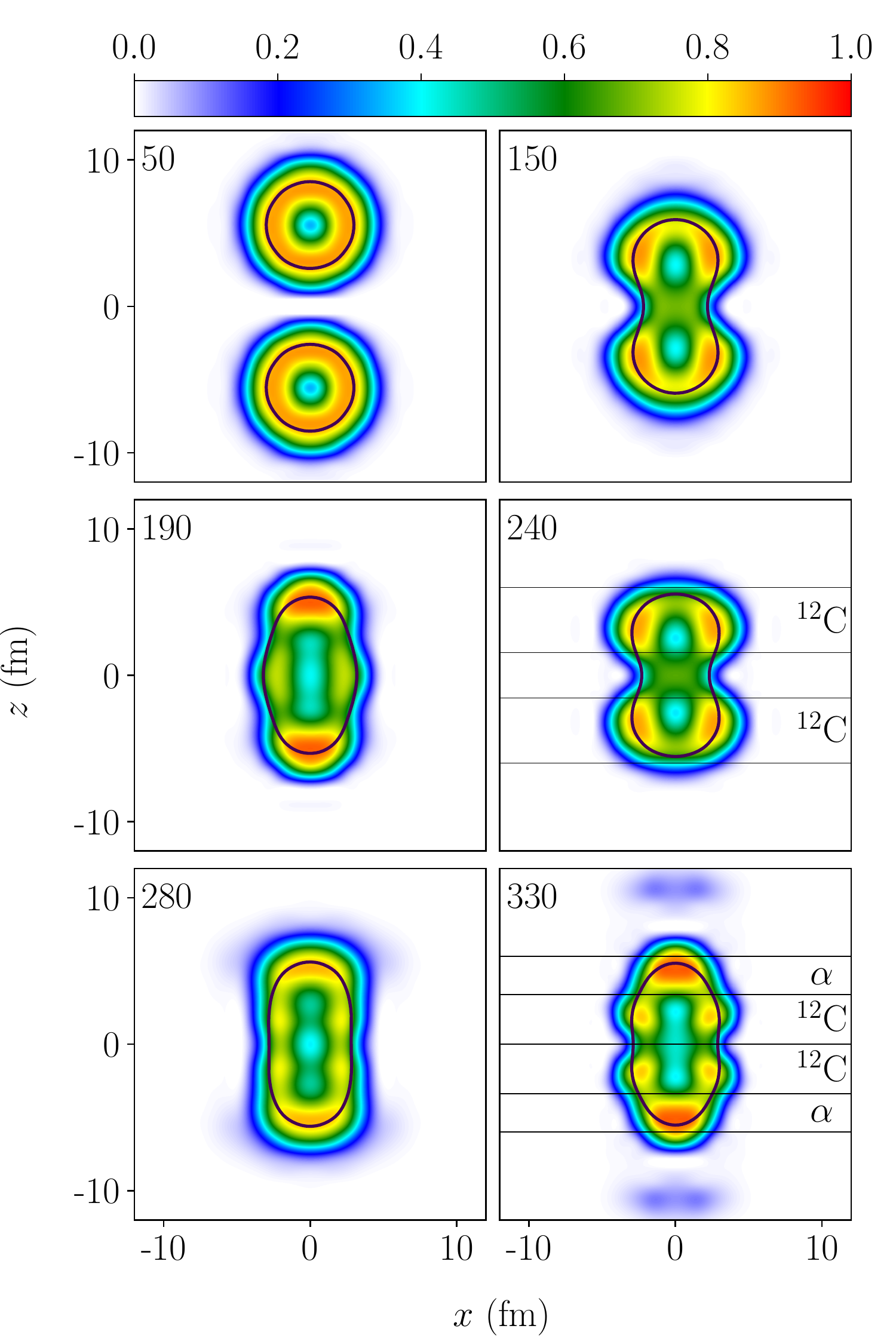}
\caption{\label{fig:OO_20_0} Localization \loc[\alpha] for the central collision of \Ox{} + \Ox{} at $\Ecm=20\MeV$. The numbers indicate the collision time (in fm/c). The black line marks the $\rho=0.05\fmd$ contour of the total density. See Supplemental Material \cite{Supplemental} for animations.}
\end{figure}
We begin  with the case of the symmetric central collision of two \Ox{} nuclei with energy  $\Ecm=20\MeV$ just above the barrier. As seen  in the $t=50\fmt$ panel of Fig.~\ref{fig:OO_20_0}, the g.s. localization of \Ox{} exhibits characteristic pattern of concentric rings, which can be associated with the filling of 0$s$ and $0p$ shells.

As the fragments come closer, the magnitude of the NLF of the fragments facing each other gets reduced, because the outer parts of the wave functions of the fragments overlap. At $t=150\fmt$ a pre-compound nucleus is formed.  At later times, the system  reveals strong \al{} clustering. As shown in the supplemental material, the pre-compound nucleus oscillates predominantly between the  structures shown in the  $t=240\fmt$ and $t=330\fmt$ panels, going through the intermediate states  displayed in  $t=190\fmt$ and $t=280\fmt$ panels. 

The configuration at $t=240\fmt$ exhibits two rings of enhanced localization at $z=\pm4\fm$. As already mentioned in Ref.~\cite{Reinhard2011}, these rings can be interpreted as oblate-deformed \Carb{} clusters.  Indeed, the nucleon content  (\ref{eq:integ_dens}) corresponding to the regions marked by horizontal lines  in Fig.~\ref{fig:OO_20_0} matches nicely the NLF ring structure. 
The central region between the two \Carb{} clusters
contains  4 neutrons and 4 protons. The structure at $t=330\fmt$  exhibits large localization at the tips, which is indicative of \al{} clustering. The interior is made of two ring structures, which we interpret  as \Carb{} oblate clusters; the nucleon content is consistent with this interpretation. 
We can thus view the  pre-compound state  depicted in  Fig.~\ref{fig:OO_20_0} as a collective oscillation of two   \Carb{} rings against two \al{} clusters.

\begin{figure}[htb]
\includegraphics[width=.9\columnwidth]{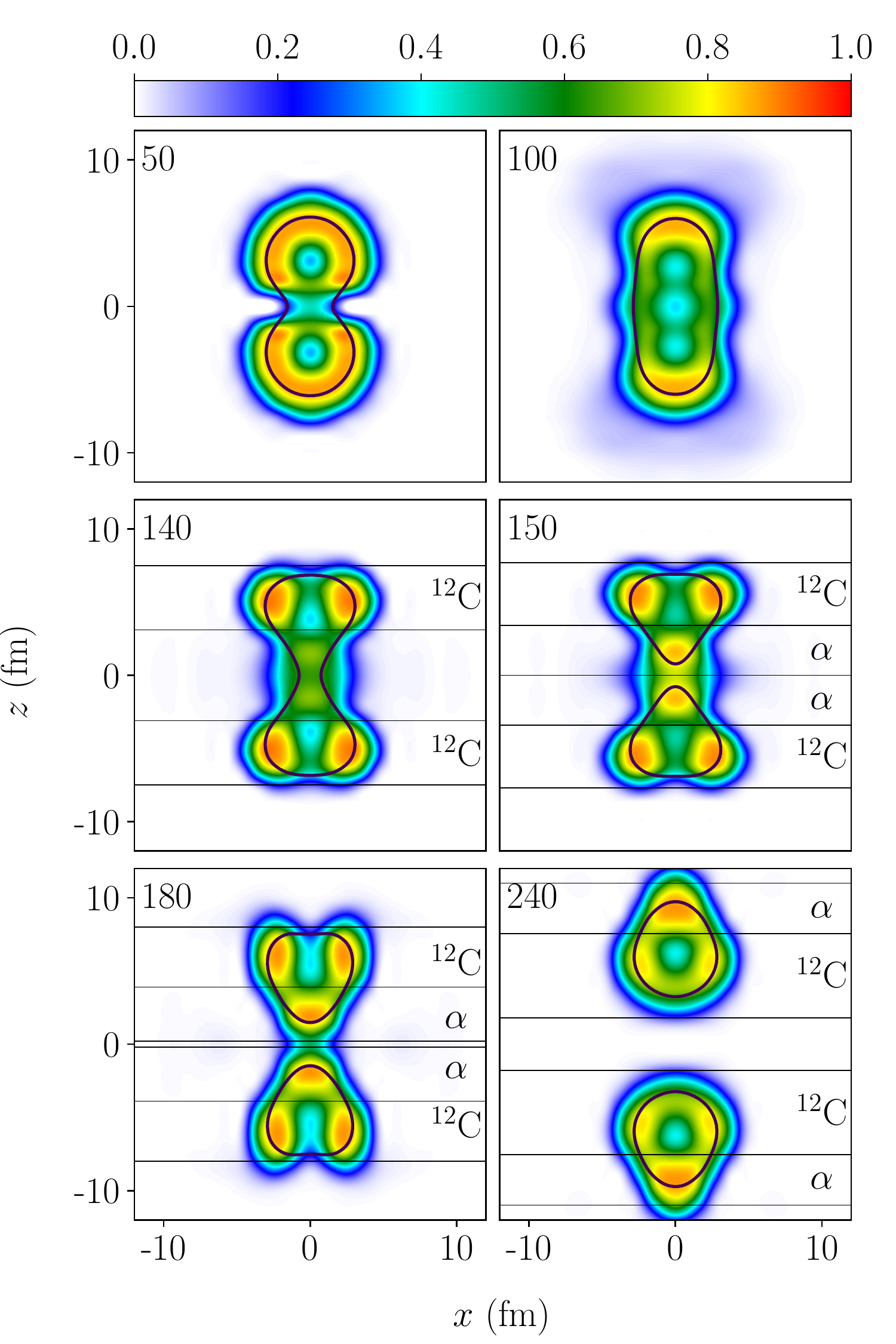}
\caption{\label{fig:OO_100_0} Similar  to  Fig.~\ref{fig:OO_20_0} except for $\Ecm=100\MeV$.}
\end{figure}
At higher center-of-mass energies, the system is expected to fission into two symmetric fragments following a brief intermediate phase.  An example
of such fusion-fission (or quasi-fission)  reaction is shown in Fig.~\ref{fig:OO_100_0}, which illustrates the \Ox{} + \Ox{} collision at  $\Ecm=100\MeV$. Following the initial contact ($t=50\fmt$), the intermediate state  is formed that eventually splits up  at $t=240\fmt$. In the intermediate state,  two \Carb{} clusters are  visible at the tips and the \al{} clusters are formed in the neck area.  Following fission, the highly excited \Ox{}  nuclei undergo octupole vibrations, in which the  \al{} cluster  oscillates with respect to the \Carb{} cluster.

\begin{figure}[tb]
\includegraphics[width=0.8\columnwidth]{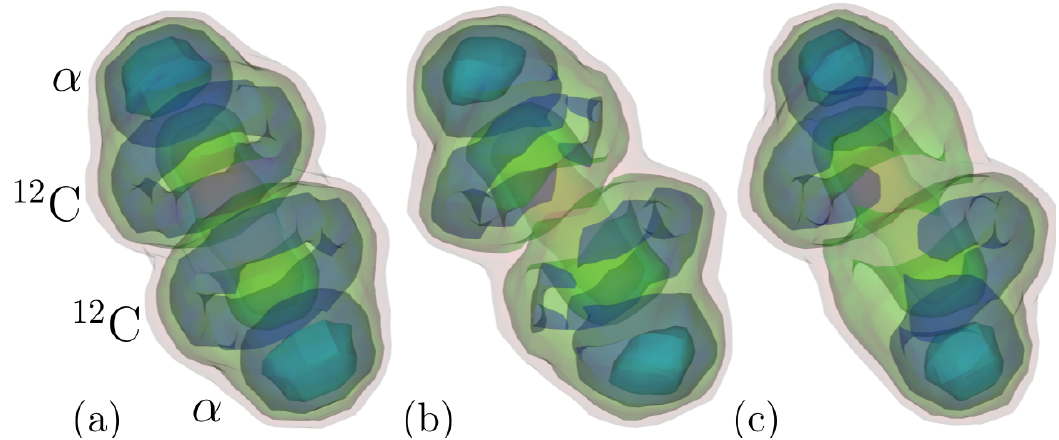}
\caption{\label{fig:aCCa} The \al{}-\Carb{}-\Carb{}-\al{} structure formed at $t=330\fmt$  in \Ox{} + \Ox{} collision at $\Ecm=20\MeV$ for three values of the impact parameter:  $b=0\fm$ (a), $b=2\fm$ (b), and  $b=4\fm$ (c). The color scale is: 0.55 light red; 0.65 green; 0.75 blue; and 0.85 cyan.}
\end{figure}

In peripheral collisions with a non-zero impact parameter  different clusters  are also predicted. Figure~\ref{fig:aCCa} shows the NLF contour plots in 3D for the \al{}-\Carb-\Carb-\al{} molecular state  found in Fig.~\ref{fig:OO_20_0} at $t=330\fmt$ for three values of the impact parameter. While for the central collision the system conserves axial symmetry, for $b>0$ the \al{} clusters shift slightly into the direction of rotation thus creating more overlap between 
\al{}  and \Carb{} clusters. A similar situation is expected for other 
configurations. 

\subsection{\Ca{} + \Ca{} collisions}

The pre-compound, or pre-fission, states produced in \Ox{} + \Ox{} collisions are expected to have a fairly simple cluster makeup. This is not going to be the case as one moves up in mass to heavier projectiles and targets.
A case in point is the   collision  of doubly-magic \Ca{} nuclei.
\begin{figure}[htb]
\includegraphics[width=.9\columnwidth]{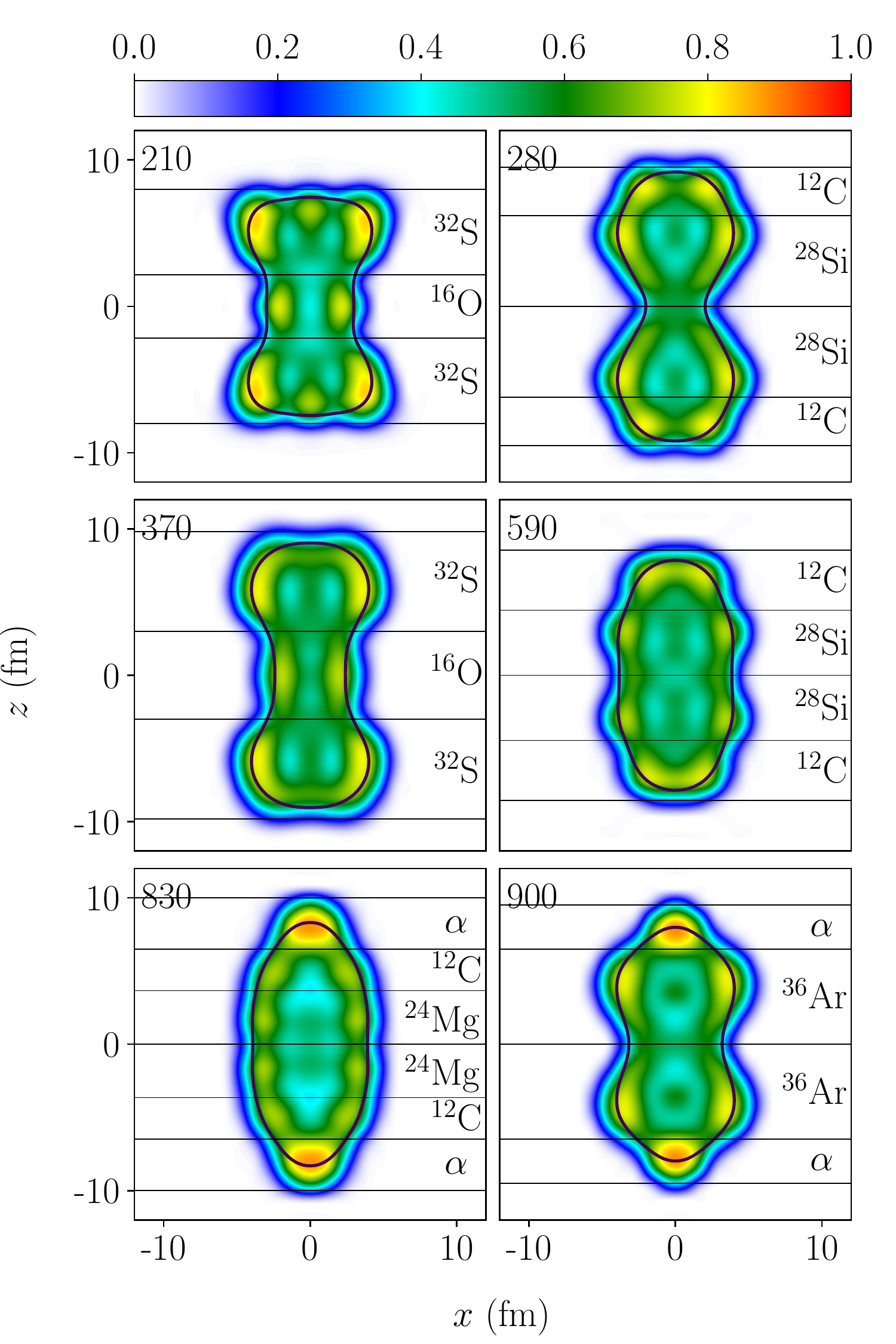}
\caption{\label{fig:CaCa_150_0} Similar  to  Fig.~\ref{fig:OO_20_0} except for the central collision of \Ca{} + \Ca{} at $\Ecm=150\MeV$. See Supplemental Material \cite{Supplemental} for animations.}
\end{figure}
Figure~\ref{fig:CaCa_150_0} shows the results of TDDFT simulations for 
the pre-compound state formed in the \Ca{}+\Ca{} central collision at at $\Ecm=150\MeV$.  In contrast to the simple \Ox{} + \Ox{} case,  the resulting excited configuration of $^{80}$Zr exhibits a rather intricate structure involving a  variety of clusters and shapes as time evolves. While in the \Ox{} collision the \al{} and \Carb{} clusters can be clearly identified through NLFs, this does not hold in general for the heavier case. 

At the early times, $t=210\fmt$ and $t=370\fmt$, the pre-compound state  can be associated with configurations involving two   $^{32}$S clusters separated by a smaller inner cluster of \Ox{}. The shapes at $t=280\fmt$ and $t=590\fmt$ consist of two smaller \Carb{} clusters at the tips and two $^{28}$Si  clusters in the interior. At the later times, however, the picture changes as a pronounced \al{} clustering appears  at the tips. At $t=830\fmt$, four rings of enhanced localization are visible within the nuclear volume. The outer rings have the \Carb{} content while the inner ones can be associated with  $^{24}$Mg. At $t=900\fmt$   the system resembles \al{}-$^{36}$Ar-$^{36}$Ar-\al{} molecular state.  As shown in the Supplemental Material \cite{Supplemental},  at $t>1000\fmt$ the the system remains in a superdeformed  shape, with the inner cluster structures evolving continuously.

\begin{figure}[htb]
\includegraphics[width=.9\columnwidth]{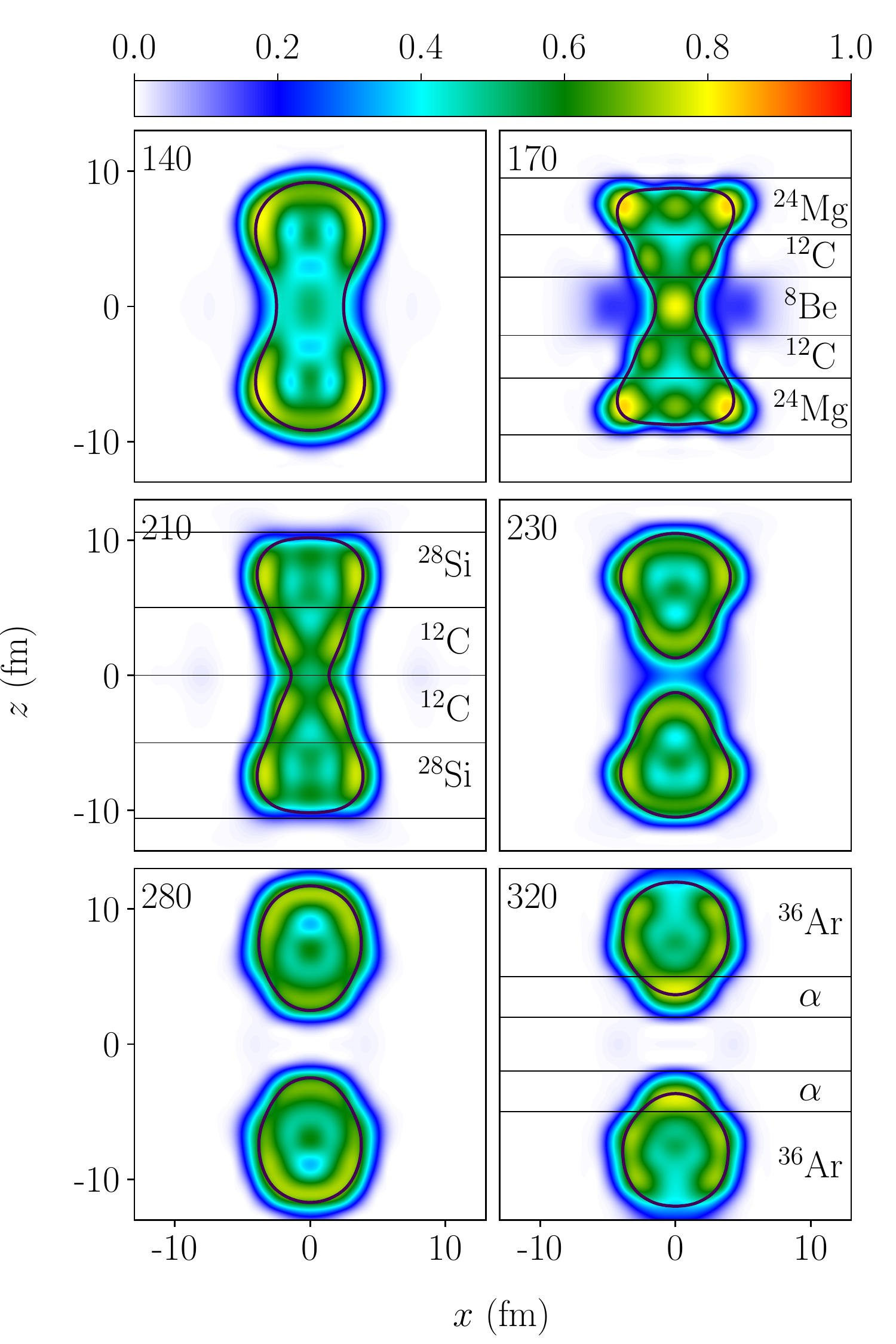}
\caption{\label{fig:CaCa_300_0} 
Similar  to  Fig.~\ref{fig:CaCa_150_0} except for $\Ecm=300\MeV$.}
\end{figure}

In Fig.~\ref{fig:CaCa_300_0} we show the fusion-fission of  \Ca{} + \Ca{} at $\Ecm=300\MeV$. Here, the intermediate state survives for only a very short time before the system splits up.  At $t=170\fmt$, oblate $^{24}$Mg clusters are visible at the tips. They are separated by two \Carb{} clusters and the region of enhanced localization in the center associated with  $^8$Be. The enhanced localization in the center vanishes at $t=210\fmt$ and only $^{28}$Si and \Carb{} cluster structures remain. After the break up, the fragments 
undergo parity-breaking oscillations along the $z$-direction. As seen in the $t=320\fmt$ panel, this octupole mode can be viewed as a vibration of the $^{36}$Ar-\al{} quasi-molecule.

\begin{figure}[htb]
\includegraphics[width=0.8\columnwidth]{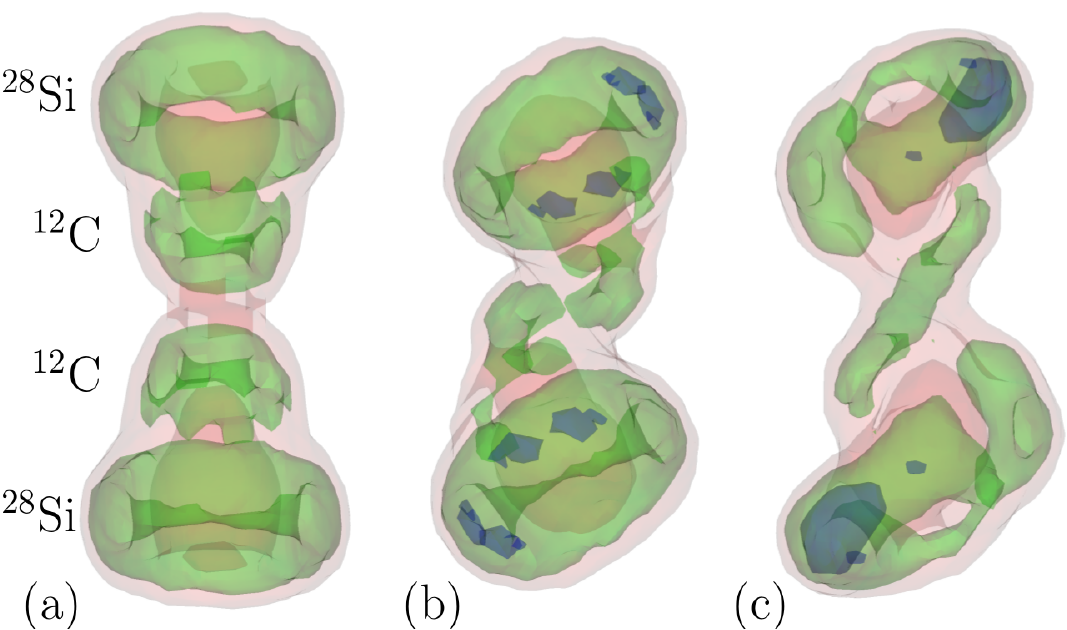}
\caption{\label{fig:Ca40b} NLF  in collision of \Ca{} + \Ca{} at $\Ecm=300\MeV$ for three values of the impact parameter:  $b=0\fm$ at $t=210\fmt$ (a); $b=3\fm$ at $t=210\fmt$ (b); and $b=6\fm$ at $t=460\fmt$ (c).}
\end{figure}
To complete the discussion,  in Fig.~\ref{fig:Ca40b} we show the NLFs for the peripheral \Ca{} + \Ca{} collision at $\Ecm=300\MeV$ just before the system's breakup. The situation resembles the results for \Ox{} + \Ox{} in Fig.~\ref{fig:aCCa}. Namely,  the rings of enhanced localization become tilted and partly overlap
at increasing values of the impact parameter.

\section{Localization in asymmetric collisions}  \label{sec:asym}
\subsection{\Ox[16]{} + \Ca{}}

For asymmetric collisions, the pre-compound state is reflection-asymmetric and its cluster content becomes fairly complex.
\begin{figure}[htb]
\includegraphics[width=.9\columnwidth]{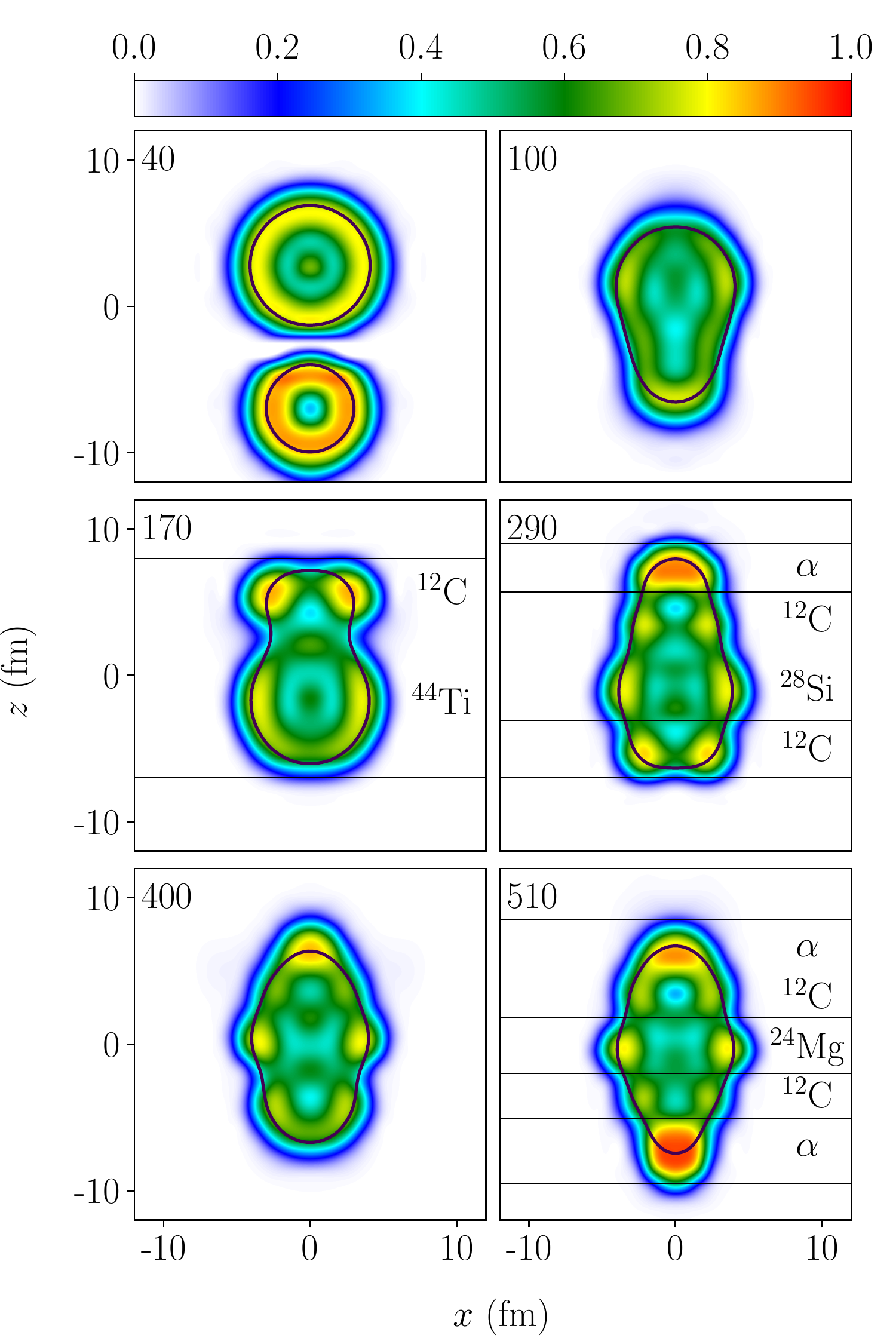}
\caption{\label{fig:OCa_80_0} Similar  to  Fig.~\ref{fig:OO_20_0} except for the central collision of \Ox{} + \Ca{} at $\Ecm=80\MeV$. See Supplemental Material \cite{Supplemental} for animations}
\end{figure}
The NLFs for the central  \Ox{} + \Ca{} collision at $\Ecm=80\MeV$ are shown in Fig.~\ref{fig:OCa_80_0}. 
A pre-compound state is formed at $t=100\fmt$. At $t=170\fmt$, a quasi-molecular
\Carb{}-$^{44}$Ti structure is predicted. At the later times, the system undergoes large-amplitude vibrations  involving different quasi-molecular configurations with oblate $^{28}$Si   and $^{24}$Mg clusters as well as intermediate states that do not exhibit a compelling cluster structure.

\begin{figure}[htb]
\includegraphics[width=0.8\columnwidth]{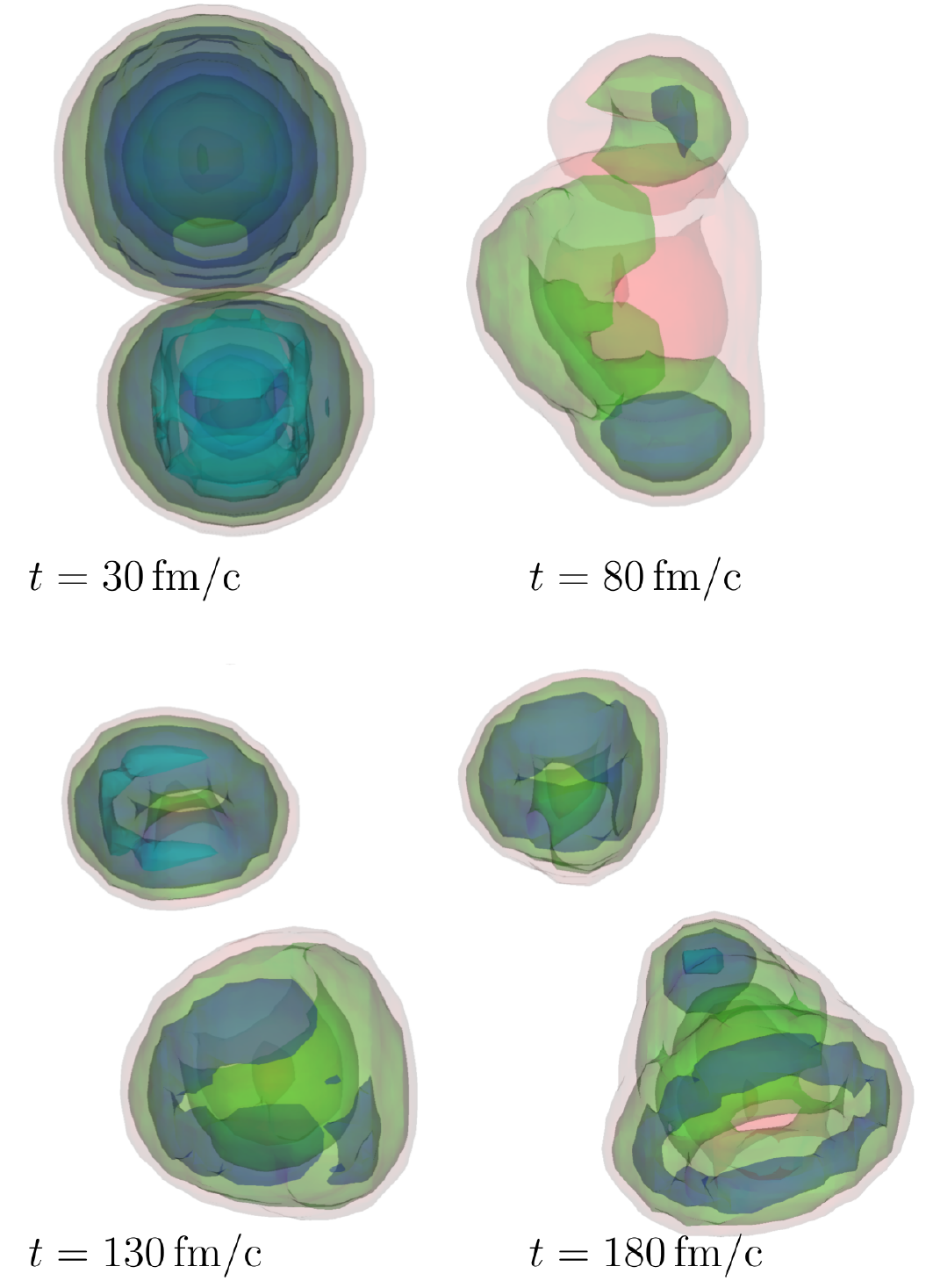}
\caption{\label{fig:OCa_200_2} 
NLF  \loc[\alpha] for the   \Ox{} + \Ca{} collision of  at $\Ecm=200\MeV$ and $b=2\fm$ at different times, as indicated.}
\end{figure}
An interesting case is the  \Ox{} + \Ca{}  collision at $\Ecm=200\MeV$ with an impact parameter of $b=2\fm$. Due to the asymmetry of the collision, the final fragments have different number of neutrons and protons. In this case, shown in Fig.~\ref{fig:OCa_200_2},  the composite system formed at $t=80\fmt$ splits up after approximately $t=150\fmt$. The mass number of the lighter fragment is $A\approx 13.7$ and its charge number is $Z\approx 7$. The snapshots at $t=130\fmt$ and $t=180\fmt$ indicate a contribution of \Carb{} cluster in the lighter fragment.

\subsection{\Ox[16,18]{} + \Carb{}}

\begin{figure}[htb]
\includegraphics[width=.9\columnwidth]{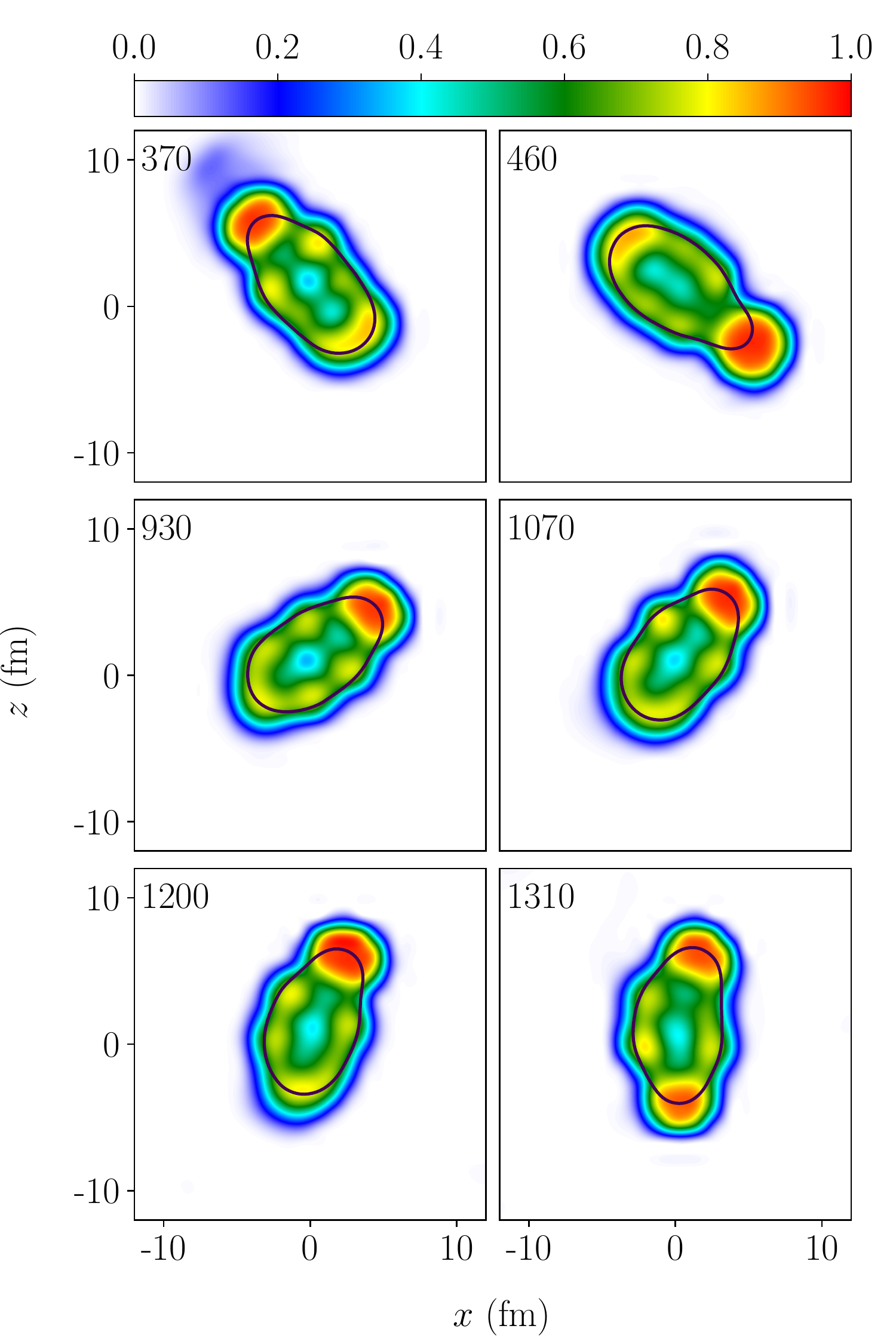}
\caption{\label{fig:O18C12_14_2} 
 Similar  to  Fig.~\ref{fig:OO_20_0} except for the \Ox[18]{} + \Carb{}  collision at $\Ecm=14\MeV$ and $b=2\fm$. See Supplemental Material \cite{Supplemental} for animations.}
\end{figure}
The  \Ox{}+\Carb{}, \Ox{}+\Carb[13]{},  and \Ox[18]{}+\Carb{} collisions have been studied experimentally in Refs.~\cite{Vadas15,CHRISTENSEN1977,Tabor77,Papadopoulos86}. In particular, in Ref.~\cite{Vadas15} the cross sections for \al{} decays following fusion at energies around the fission barrier ($6-14\MeV$) have been studied. The authors concluded that statistical models strongly underestimate  \al{} emission. A possible reason is that  statistical models do not take into account the entrance channel effects in the pre-compound system, but assume it to be completely thermalized. In this work, we cannot estimate \al{} emission effects, because TDDFT is unable to  describe quantum tunneling. However, with the help of the localizations $\loc[\alpha]$ we can assess  the formation of \al{}-clusters effects in the pre-compound system. 

Figure~\ref{fig:O18C12_14_2} shows the  snapshots of \loc[\alpha] in the \Ox[18]{}+\Carb{} reaction  at $\Ecm=14\MeV$ and $b=2\fm$. Appreciable \al{} clustering effects are apparent,  especially at the tips of the pre-compound system. Our calculations indicate that the tendency for \al{} clusters to appear is strong for  energies between 8 and $14\MeV$ and the impact parameters for which the system fuses. The collision of \Ox{}+\Carb{} reveals a very similar behavior. 
While our TDDFT calculations can  shed light on the  formation process of \al{} clusters, we cannot 
directly address the experimental data  for  the \al{} decay cross section and \al{} emission probabilities. Such a task would require significant extensions of the current framework.

\section{Conclusions} \label{sec:Conclusion}

We used the time-dependent localization functions \loc[\alpha] to illustrate cluster effects in 
TDDFT simulations of the low-energy heavy-ion collisions. Compared to the particle density $\rho$ , the localization  \loc[\alpha] provides excellent measure of clusters of \al{} particles and \al-conjugate nuclei appearing  in the  pre-compound, or pre-fission,  states produced in nuclear collisions. In this context, Video 2 in the Supplemental Material \cite{Supplemental} nicely illustrates the advantage of using  \loc[\alpha]  over $\rho$.

In the central \Ox{} + \Ox{} collision, \al{} and \Carb{} clusters are predicted to be  formed. In  reactions involving \Ca{}, heavier 
clusters of $\alpha$-conjugate nuclei are also expected. Moreover, our analysis indicates that the large amplitude collective motion  of the pre-compound system is far more complex than what is suggested by  a na\"{\i}ve  liquid drop picture of vibrating nucleonic fluids. Namely, in TDDFT, the resulting collective mode involves  
cluster motion  within quasi-molecular configurations, as well as exchange of \al{} particles between clusters, leading to cluster transmutations  in  heavier systems.  Of special interest are the fusion-fission  reactions at higher energies, where strong clustering phenomena are predicted both before and after   breakup.
 
For the collision of \Ox[16,18]{} + \Carb{} we showed that the pre-compound system has strong tendency to form \al{} clusters. This result
supports the conclusions of Ref.~\cite{Vadas15} that the 
cluster structure of the initial projectile and target nuclei gives rise to
strong entrance channel effects and influences the 
\al{} emission following fusion. In order to estimate the actual
pre-equilibrium 
\al{}-emission probability, significant extensions of the formalism by going beyond TDDFT are required. Work along such lines is in progress.


\begin{acknowledgements}
 This work was supported by the U.S. Department of Energy under Award Numbers DOE-DE-NA0002847 (NNSA, the Stewardship Science Academic Alliances program), DE-SC0013365 (Office of Science), DE-SC0008511 (Office of Science, NUCLEI SciDAC-3 collaboration) and BMBF-Verbundforschungsprojekt number 05P15RDFN1. An award of computer time was provided by the Institute for Cyber-Enabled Research at Michigan State University. 
\end{acknowledgements}
\bibliographystyle{apsrev4-1}
\bibliography{localization}

\end{document}